\documentclass[letterpaper, 10 pt, conference]{ieeeconf}  

\IEEEoverridecommandlockouts                              
\usepackage{graphics}
\usepackage{amsmath}
\usepackage{amssymb}
\usepackage{cite}

\usepackage{graphicx}
\newcommand{\argmax}{\mathop{\rm arg\ max}\limits}

\title{\LARGE \bf
Recognition of Unseen Combined Motions via Convex Combination-based EMG Pattern Synthesis for Myoelectric Control
}

\author{Itsuki Yazawa$^{1}$, Seitaro Yoneda$^{1}$ and Akira Furui$^{1}$
\thanks{This work was partially supported by JSPS KAKENHI Grant Number JP23K28128.}
\thanks{$^{1}$I. Yazawa, S. Yoneda and A. Furui are with Graduate School of Advanced Science and Engineering,
        Hiroshima University, Higashi-hiroshima, Japan 
        (e-mail: {\tt\small \{itsukiyazawa, seitaroyoneda, akirafurui\}@hiroshima-u.ac.jp})}%
}

\begin{document}

\maketitle
\thispagestyle{empty}
\pagestyle{empty}

\begin{abstract}
Electromyogram (EMG) signals recorded from the skin surface enable intuitive control of assistive devices such as prosthetic limbs.
However, in EMG-based motion recognition, collecting comprehensive training data for all target motions remains challenging, particularly for complex combined motions.
This paper proposes a method to efficiently recognize combined motions using synthetic EMG data generated through convex combinations of basic motion patterns.
Instead of measuring all possible combined motions, the proposed method utilizes measured basic motion data along with synthetically combined motion data for training.
This approach expands the range of recognizable combined motions while minimizing the required training data collection.
We evaluated the effectiveness of the proposed method through an upper limb motion classification experiment with eight subjects.
The experimental results demonstrated that the proposed method improved the classification accuracy for unseen combined motions by approximately 17\%.
\end{abstract}

\section{Introduction}

Electromyogram (EMG) signals are biosignals that can be noninvasively measured from the skin surface, reflecting action potentials associated with muscle contractions.
These signals directly represent human movement intentions, making them widely studied as intuitive interfaces for controlling assistive devices, particularly prosthetic limbs~\cite{geethanjali2016myoelectric, zecca2002control, al2024using, siguencia2020development}.
Recent advancements in machine learning-based classification models have enabled increasingly accurate estimation of user motion intentions from measured EMG signals.

The implementation of EMG-based motion classification models typically requires collecting training data for all target motions.
As the number of target motions increases, the volume of required training data grows proportionally, creating a significant burden on users.
This challenge is particularly pronounced for human upper limb motions, which involve complex multi-degree-of-freedom (multi-DoF) motions.
Consequently, it becomes impractical to comprehensively collect training data for all possible motions due to user fatigue and time constraints.
Therefore, current myoelectric prosthesis systems are often limited to classifying only a restricted set of motions~\cite{atzori_deep_2016, atzori2014electromyography, saito2022evidence}.

To address this limitation, researchers have made various efforts to efficiently solve the multi-DoF motion classification problem by representing target motions as combinations of fundamental motion components~\cite{shima_classification_2010, krasoulis2020myoelectric, olsson2019extraction, zbinden2024sequential}.
Notably, Shima~\textit{et al.} proposed an approach based on the concept of muscle synergies, in which the outputs of a classification model trained solely on basic motion data are combined to recognize previously unseen combined motions~\cite{shima_classification_2010}.
Their approach classified combined motions by calculating their similarity to predefined probability vectors.
This method enables the classification of multi-DoF motions by simply modifying these probability vectors, without requiring additional training data or changes to the model architecture.
However, models trained exclusively on basic motion data may struggle in effectively distinguish between basic and combined motions due to the lack of combined motion patterns in the training.

In this paper, we propose a method for more robust recognition of unseen combined motions through the generation and utilization of synthetic EMG data for training classification models.
We define \textit{basic motions} as individual, elementary movements (e.g., wrist flexion, hand opening), and \textit{combined motions} as concurrent executions of multiple basic motions (e.g., simultaneous wrist flexion and hand opening).
The proposed method generates synthetic EMG patterns for combined motions through convex combinations of basic motion patterns, enabling the model to learn more distinctive features for both motion types.
This approach reduces the required number of measured motions while enabling robust recognition of diverse combined motions, thereby overcoming the limitations of previous approaches that rely solely on basic motion data.

\section{Proposed Method}

This section presents a classification method for EMG signals recorded from $D$ electrodes, which categorizes signals into $M_b$ basic motion classes and $M_c$ combined motion classes.
Each combined motion class is defined as a combination of two or more basic motion classes.
The proposed method requires only EMG patterns from basic motion classes during the training phase.
These patterns are then utilized to synthetically generate EMG patterns corresponding to combined motion classes, which serve as training data for the classification model.
Our method employs a neural network as the classification model, where synthetic EMG patterns are generated by combining outputs from any layer of the network, including the input layer.

\subsection{Synthetic Data Generation and Model Training}

\begin{figure*}[t]
	\centering
	\includegraphics[width=0.7\hsize]{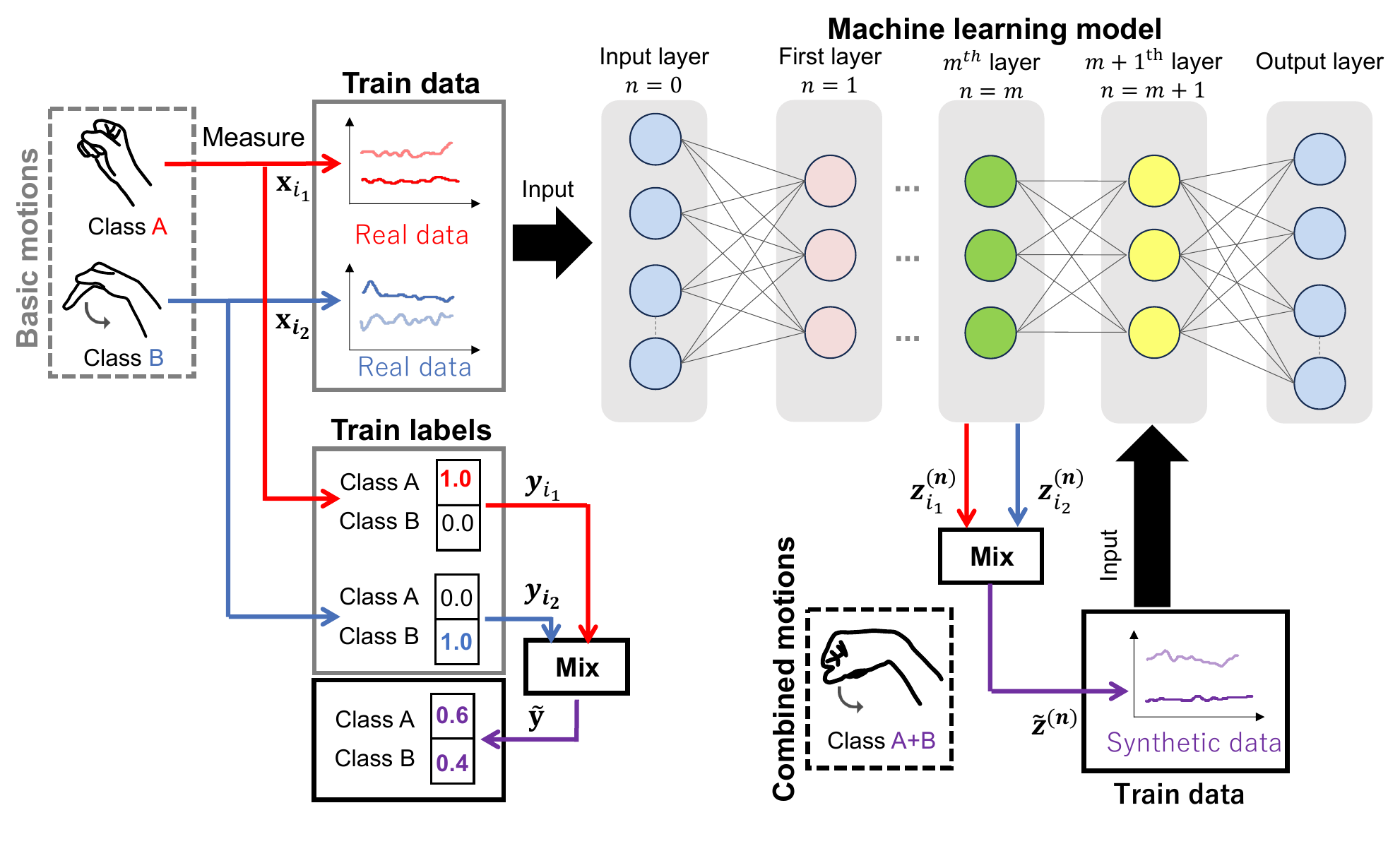}
	\caption{Schematic overview of the proposed method for generating synthetic EMG patterns of combined motions
    using convex combinations of basic motion patterns at different neural network layers}
	\label{overview}
\end{figure*}

Fig.~\ref{overview} shows the training phase of the proposed method.
The synthetic data generation process utilizes a basic motion dataset $\mathcal{D}_b=\{(\mathbf{x}_i,\mathbf{y}_i)\}_{i=1}^{N_b}$, where $\mathbf{x}_i \in \mathbb{R}^D$ represents an EMG pattern and $\mathbf{y}_i \in \{0,1\}^{M_b}$ denotes its corresponding one-hot encoded class label.

For the neural network classifier $f$, let $g_n$ denote the mapping from the input layer to the $n$th layer, and $f_n$ represent the mapping from the $n$th layer to the output layer.
The complete model can be expressed as $f(\mathbf{x})=f_n(g_n(\mathbf{x}))$.

To generate synthetic data for combined motion class $m$, we selectively extract data from $K_m$ distinct basic motion classes.
The synthetic data and their corresponding labels are generated using convex combinations of the representations at the $n$th layer of the neural network~\cite{guo2020nonlinear}.
To maintain consistency between feature and label spaces, both synthetic features $\tilde{\mathbf{z}}^{(n)}$ and their corresponding labels $\tilde{\mathbf{y}}$ are generated using identical coefficients $\lambda_k$ as follows:
\begin{align}
    \label{mix_data}
    \tilde{\mathbf{z}}^{(n)} &= \sum_{k=1}^{K_m} \lambda_k g_n(\mathbf{x}_{i_k}), \\
    \label{mix_label}
    \tilde{\mathbf{y}} &= \sum_{k=1}^{K_m} \lambda_k \mathbf{y}_{i_k}.
\end{align}
When $n=0$, synthesis occurs at the input layer, where $g_0$ is the identity mapping ($g_0(\mathbf{x})=\mathbf{x}$), indicating direct combination in the input space.
Conversely, when $n\ge1$, synthesis occurs at the middle layer.
In this condition, input data are transformed to generate synthetic data in the nonlinear space.
The coefficients $\lambda_k$ represent mixing ratios constrained by $\lambda_k \in [0,1]$ and $\sum_k \lambda_k=1$.

To generate diverse yet controlled mixing ratios that satisfy these constraints, we sample $\lambda_k$ from a symmetric Dirichlet distribution:
\begin{align}
	\label{dirichlet}
	p(\lambda_k \vert \alpha) &= \frac{1}{B(\mathbf{\alpha})} \prod_{k=1}^{K_m} \mathbf{\lambda}_{k}^{\alpha-1}, \\
	B(\mathbf{\alpha}) &= \frac{\Gamma(\alpha)^{K_m}}{\Gamma(K_m \alpha)}.
\end{align}
The parameter $\alpha>0$ serves as a concentration parameter controlling the distribution shape: larger values of $\alpha$ result in mixing ratios that are more uniformly distributed across components.
This sampling scheme enables the generation of diverse synthetic data with varying mixing ratios while maintaining the relationship between features and labels~\cite{zhang_mixup_2017}.

We construct the final training dataset $\mathcal{D}$ by combining the basic motion dataset $\mathcal{D}_b$ with the synthetic combined motion dataset $\tilde{\mathcal{D}}_c=\{(\tilde{\mathbf{z}}^{(n)}_j,\tilde{\mathbf{y}}_j)\}^{N_c}_{j=1}$, such that $\mathcal{D}=\mathcal{D}_b\cup\tilde{\mathcal{D}}_c$.
This unified dataset is used to train the classification model $f$ by optimizing the following composite loss function:
\begin{align}
    \mathcal{L}=\frac{1}{N_b}\sum^{N_b}_{i=1}\ell(f(\mathbf{x}_i), \mathbf{y}_i) + \frac{1}{N_c}\sum^{N_c}_{j=1}\ell(f_n(\tilde{\mathbf{z}}^{(n)}_j), \tilde{\mathbf{y}}_j).
    \label{loss_function}
\end{align}
The loss function comprises two components: The first term, $\ell(f(\mathbf{x}_i),\mathbf{y}_i)$, computes the cross-entropy loss between the model output $f(\mathbf{x}_i)$ and the corresponding target label $\mathbf{y}_i$ for basic motion patterns.
The second term computes the loss between the forward-propagated output $f_n(\tilde{\mathbf{z}}^{(n)}_j)$ of synthetic patterns (propagated from the $n$th layer) and their assigned target labels $\tilde{\mathbf{y}}_j$.
We optimize the model parameters through iterative updates that minimize this composite loss function, thereby learning to classify both basic and combined motions effectively.

\subsection{Classification in the Evaluation Phase}

In the evaluation phase, the trained model processes new test data $\mathbf{x}$ to perform classification.
The model outputs a predicted probability vector $\hat{\mathbf{y}}=[\hat{y}_1,\dots,\hat{y}_{M_b}]^{\top}=f(\mathbf{x})\in[0,1]^{M_b}$, where each element represents the probability score for one of the $M_b$ basic motion classes.
While this enables the classification of the $M_b$ basic motions, it does not directly support classification of combined motions.

\begin{figure}[t]
	\centering
	\includegraphics[width=0.7\hsize]{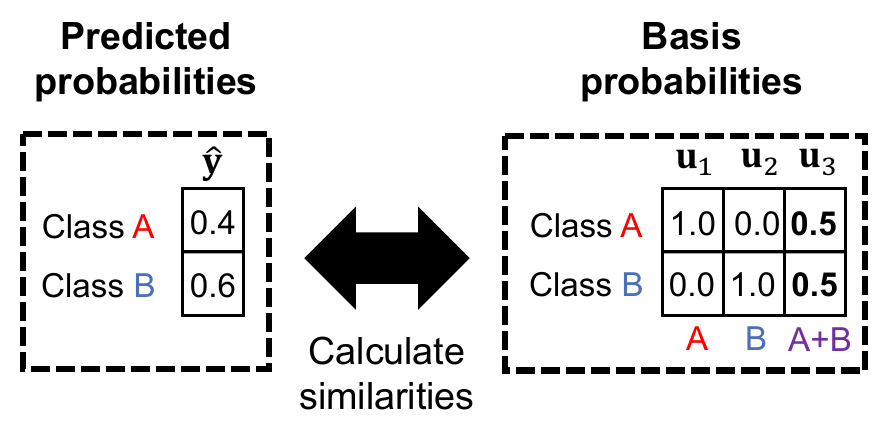}
	\caption{Illustration of similarity calculation between predicted probabilities and basis probability vectors for motion classification}
	\label{overview_similarity}
\end{figure}

To address this limitation, we introduce a similarity-based classification approach (Fig.~\ref{overview_similarity}) that enables the recognition of combined motions.
From (\ref{mix_label}), the synthetic data for combined motions are trained with soft labels, which are convex combinations of their constituent basic motion labels.
Test data from basic motions yield high probability scores for their respective motion classes, whereas those from combined motions produce simultaneously high scores for multiple constituent basic motion classes.

We thus compute the similarity $\mathcal{S}$ between the predicted probability vector $\hat{\mathbf{y}}$ and a predefined set of $M_b+M_c$ probability vectors $\mathbf{u}_m \in [0,1]^{M_b}$ ($m=1,\ldots,M_b,M_b+1,\ldots,M_b+M_c$).
The classification result $\hat{m}$ is determined by:
\begin{align}
    \hat{m} = \argmax_m \mathcal{S}(\hat{\mathbf{y}},\mathbf{u}_m),
    \label{similarity}
\end{align}
where $\mathbf{u}_m$ is a one-hot vector for basic motions ($m=1,\dots,M_b$) or a vector with values of $\frac{1}{K_m}$ for elements corresponding to the $K_m$ basic motions in combined motions ($m=M_b+1,\dots,M_b+M_c$).

We adopt the Kullback-Leibler divergence as the similarity measure $\mathcal{S}$ between the predicted and reference probability distributions:
\begin{align}
    \mathcal{S}(\hat{\mathbf{y}}, \mathbf{u}_m) = \sum_m \mathbf{u}_m \log \left(\frac{\mathbf{u}_m+\epsilon}{\hat{\mathbf{y}}+\epsilon}\right),
    \label{kl}
\end{align}
where $\epsilon$ is a small positive constant included to prevent numerical instability.

This section has presented an approach for combined motion classification which generates synthetic training data using coefficients $\lambda_k$ sampled from a symmetric Dirichlet distribution.
These coefficients enable the creation of realistic combined motion patterns through convex combinations of $K_m$ basic motion patterns.
The proposed method achieves robust classification of previously unseen combined motions through training with a combination of measured basic motion data and synthetically generated combined motion data.

\section{Experiments}

\subsection{Experimental Setup} 
We evaluated the effectiveness of the proposed method through an upper-limb motion classification experiment.
Eight healthy adults participated in the study (mean age: $23\pm0.76$ years).
Eight surface electrodes ($D=8$) were attached to the right forearm of each participant, positioned circumferentially near the elbow.
EMG signals were recorded using a wireless measurement system (Trigno, Delsys Inc., USA) at a sampling frequency of $2000$ Hz.
During the experiment, participants maintained a seated posture with the right elbow resting on a table and performed motions.
Prior to the study, participants gave their written consent after being briefed about the research objectives. 
The study protocol received approval from the Ethics Committee at Hiroshima University (approval number: E-840).

\begin{figure}[t]
	\centering
	\includegraphics[width=1\hsize]{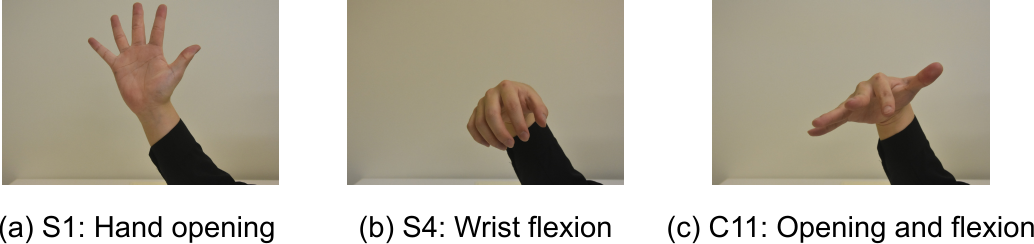}
	\caption{Examples of basic motions (S1, S4) and combined motions (C11)}
	\label{motions}
\end{figure}

Fig.~\ref{motions} shows examples of basic and combined motions performed during the experiment. 
We defined six basic motions ($M_b=6$): hand opening (S1), hand grasping (S2), wrist extension (S3), wrist flexion (S4), pronation (S5), and supination (S6). 
Additionally, we included $12$ combined motions ($M_c=12$, $K_m=2$): opening and pronation (C1), grasping and pronation (C2), extension and pronation (C3), flexion and pronation (C4), opening and supination (C5), grasping and supination (C6), extension and supination (C7), flexion and supination (C8), opening and extension (C9), grasping and extension (C10), opening and flexion (C11), and grasping and flexion (C12).
Each participant performed the complete set of motions (S1--S6 and C1--C12) in sequence, maintaining each motion for $4$ s.
This sequence was repeated across six trials.
We incorporated rest periods between motions ($4$ s) and between trials ($40$ s), during which participants maintained their posture while relaxing their muscles.

\subsection{Data Processing}
The raw EMG signals were preprocessed using a fourth-order band-stop filter (cutoff frequencies: $60$--$62$ Hz) to remove power supply noise components.
The feature extraction process consisted of full-wave rectification and subsequent smoothing with a second-order Butterworth low-pass filter (cutoff frequency: $2.0$ Hz).
To exclude the transition state from the resting posture, we removed the initial $5$\% of each data sample from the analysis.
We then implemented a two-step normalization process on the rectified and smoothed signals.
First, each electrode signal component was divided by its maximum value within the training dataset.
Second, at each data point, the signals were normalized such that the sum across all electrodes equals $1$.
These normalized EMG patterns, denoted as $\mathbf{x}$, were used as feature vectors for motion classification.

For the classification analysis, we randomly selected two trials from the basic motions as the training data, while the remaining four trials of basic motions and four trials of combined motions were used as the test data.
The average classification accuracy was calculated across all possible combinations of these trials.
Note that combined motion data corresponding to the same trials as the basic motions used in the training dataset were excluded from the analysis.

\subsection{Evaluation Methods}

The performance of the proposed method was evaluated through two main analyses: the effect of synthesis layer selection and comparison with baseline methods.
First, we investigated how the choice of synthesis layer affects classification accuracy by comparing synthesis at the input layer ($n=0$) with synthesis at each hidden layer ($n=1,2,3$) to assess the impact of feature space transformation.

We then established performance bounds by comparing our method against two baseline approaches:
\begin{itemize}
    \item Fully-supervised: This approach represented the upper bound of performance by utilizing actual combined motion data, including the two trials of combined motions not used in the proposed method.
    The classification task was treated as an $18$-class problem, encompassing all six basic motions and $12$ combined motions.
    \item Basic-only\cite{shima_classification_2010}: This approach established the lower bound by utilizing only basic motion data for training.
    Classification was performed based on similarity measures.
\end{itemize}
All methods were implemented using a three-layer multi-layer perceptron optimized with the Adam algorithm.
The hyperparameters were set as follows: a batch size of $64$, a learning rate of $1.0\times10^{-4}$, and $20$ epochs.
In the proposed method, the Dirichlet distribution parameter was set to $\alpha=50$, and synthetic data were generated in equal amounts to the measured data ($N_b=N_c$).
Based on the layer-wise comparison results, we selected the layer that achieved the highest classification accuracy for the baseline comparison.

\section{Results}
\subsection{Evaluation of Synthetic EMG Pattern Generation}
\begin{figure}[t]
	\centering
	\includegraphics[height=1\hsize]{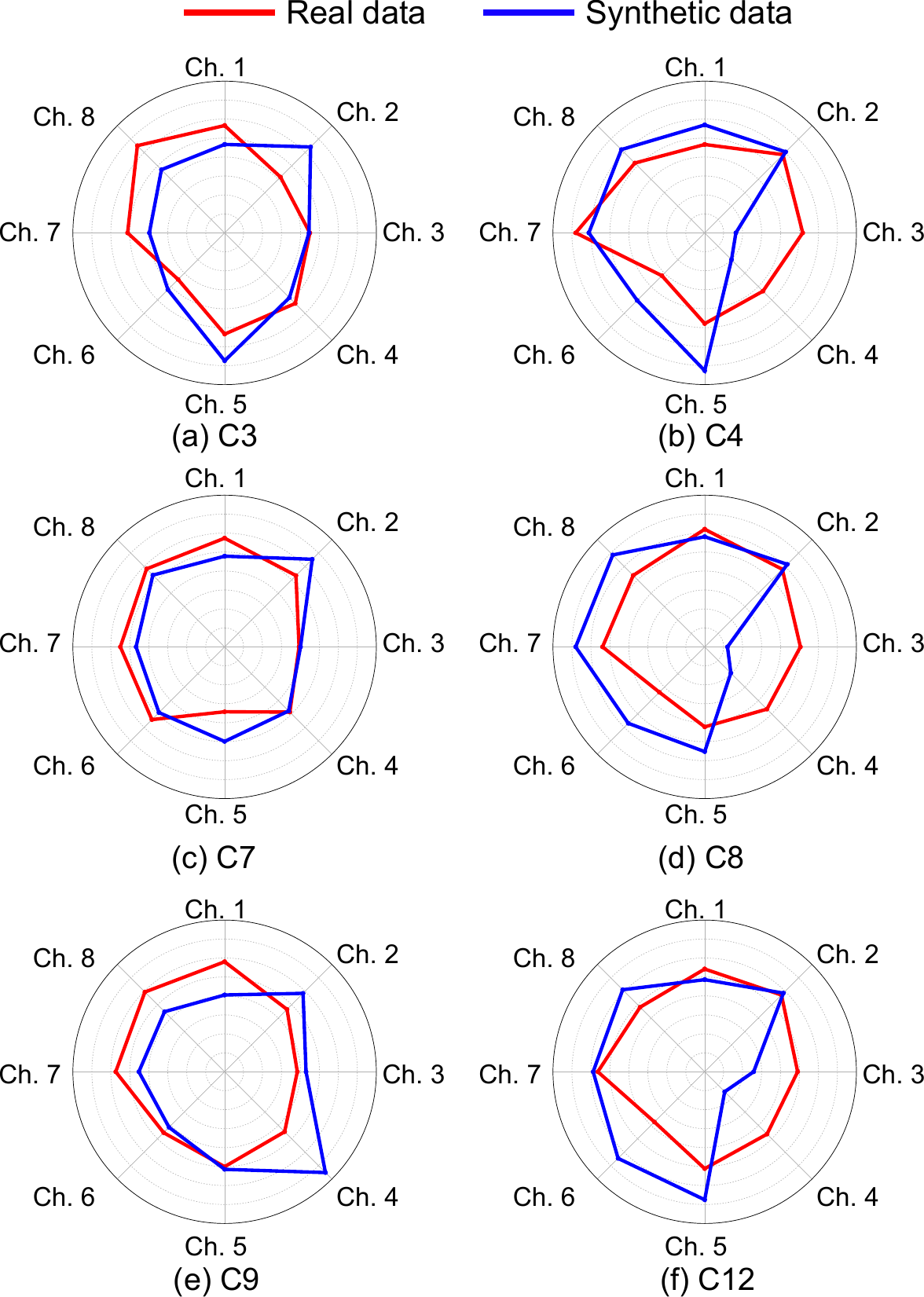}
	\caption{Comparison of actual and synthetic EMG patterns using radar plots  for six combined motions (C3, C4, C7, C8, C9, and C12)}
	\label{radarchart}
\end{figure}

Fig.~\ref{radarchart} shows a comparison between actual EMG patterns of combined motions and synthetic EMG patterns generated by the proposed method.
The radar charts display time-averaged EMG patterns for specific trials, with electrode values arranged radially.
Red patterns correspond to the measured EMG, while blue patterns represent the synthetic EMG.
Although the overall shapes demonstrate similarity, notable discrepancies in electrode values are observed in Figs.~\ref{radarchart} (b), (d), and (f).

\begin{figure}[t]
    \centering
    \includegraphics[width=0.95\hsize]{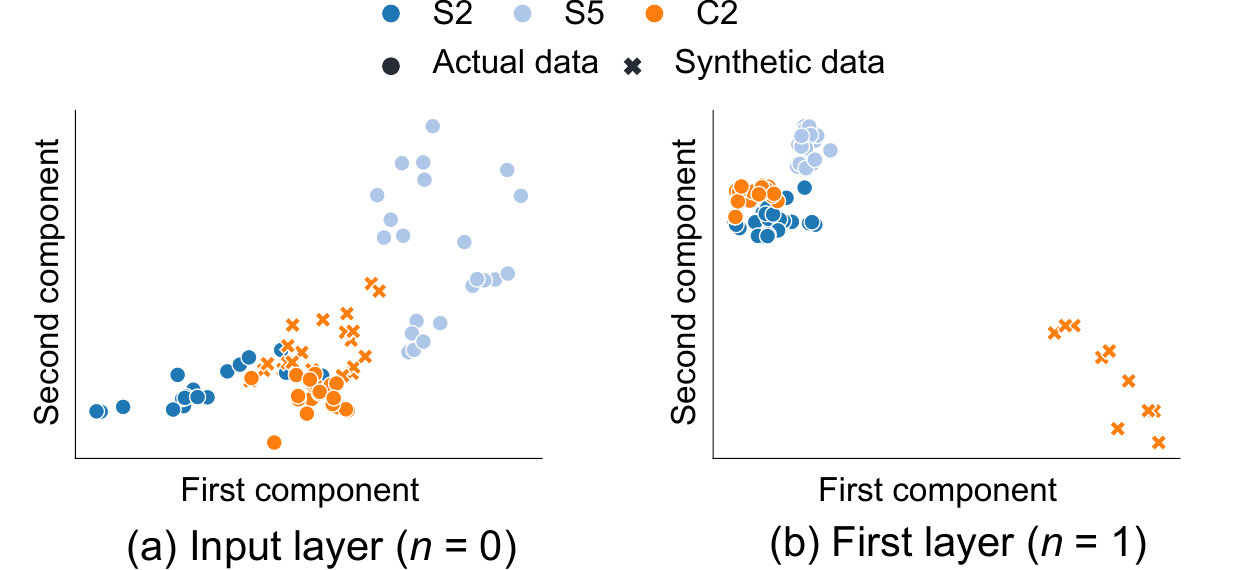}
    \caption{Examples of principal component analysis visualization comparing actual and synthetic EMG patterns at the (a) input layer and (b) first hidden layer for three motion classes (S2, S5, and C2).}
    \label{pca}
\end{figure}

Fig.~\ref{pca} illustrates the results of principal component analysis performed on both actual and synthetic EMG patterns, comparing the effectiveness of data generation at the input layer versus that at the hidden layer. 
Circles and crosses represent actual and synthetic patterns, respectively.
Blue and light blue indicate basic motions (S2, S5), and orange represents combined motion (C2).
The analysis reveals that synthetic patterns generated at the input layer closely approximate the actual patterns, whereas synthetic patterns generated at the hidden layer exhibit substantial deviation.


\subsection{Classification Performance Analysis}
\begin{figure}[t]
    \centering
    \includegraphics[width=0.7\hsize]{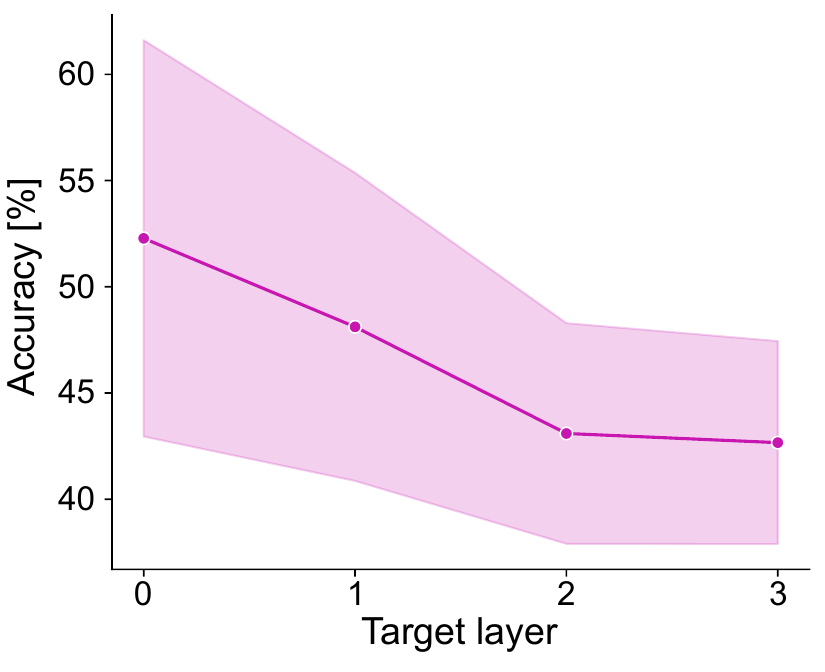}
    \caption{Average classification accuracy versus layer depth for synthetic EMG pattern generation with shaded regions indicating standard deviation}
    \label{line plot}
\end{figure}

Fig.~\ref{line plot} shows the classification accuracy across different layers of data generation.
The vertical axis represents classification accuracy, while the horizontal axis indicates the layer where data synthesis occurred.
The solid line represents mean values, with the shaded area indicating standard deviations.
The results demonstrate that synthesis at the input layer achieved the highest accuracy.
Therefore, we generate synthetic data at the input layer in the subsequent comparison.

\begin{figure}[t]
	\centering
	\includegraphics[width=1\hsize]{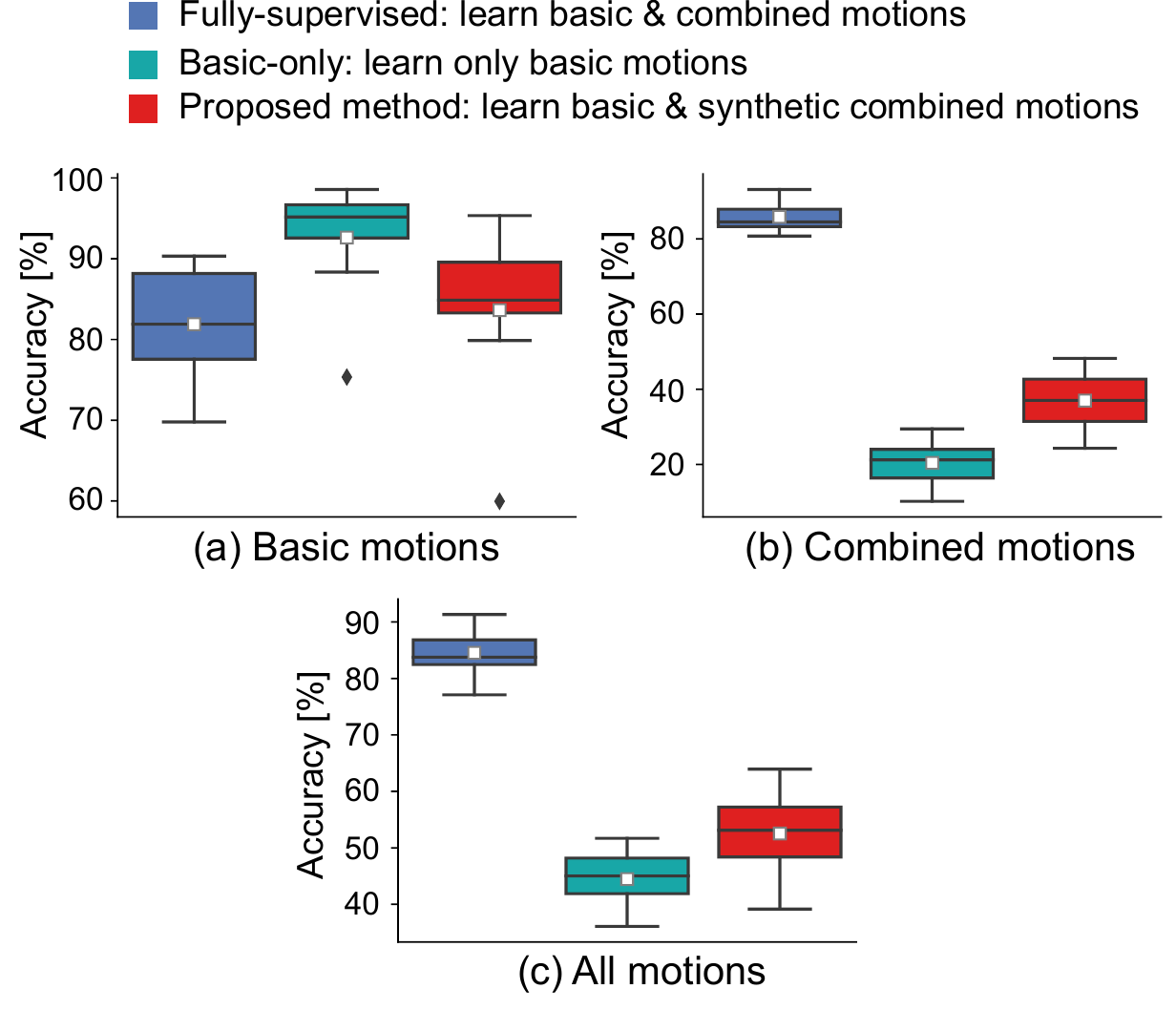}
	\caption{Classification accuracy for three learning methods on (a) basic motions, (b) combined motions, and (c) all motions. Synthesis is performed at the input layer in the proposed method.  }
	\label{box plot}
\end{figure}

Fig.~\ref{box plot} compares the classification accuracy across different methods.
Blue and light blue boxes represent the comparison methods (fully-supervised and basic-only), while red boxes indicate the results of the proposed method.
White dots denote the mean values.
The fully-supervised approach, trained with actual EMG patterns for all $18$ motions (basic and combined), achieved over $80$\% accuracy for both motion types, and demonstrated superior performance in overall motion classification accuracy compared to other methods.
This method represents an ideal scenario, where we assume complete training data availability.
The basic-only method, trained exclusively on basic motion data, achieved over $90$\% accuracy for basic motions but only $20$\% accuracy for combined motions.
In contrast, the proposed method, utilizing synthetic data generation, improved the classification accuracy of combined motions to approximately $35$\%.

\begin{figure}[t]
	\centering
        \includegraphics[width=0.75\hsize]{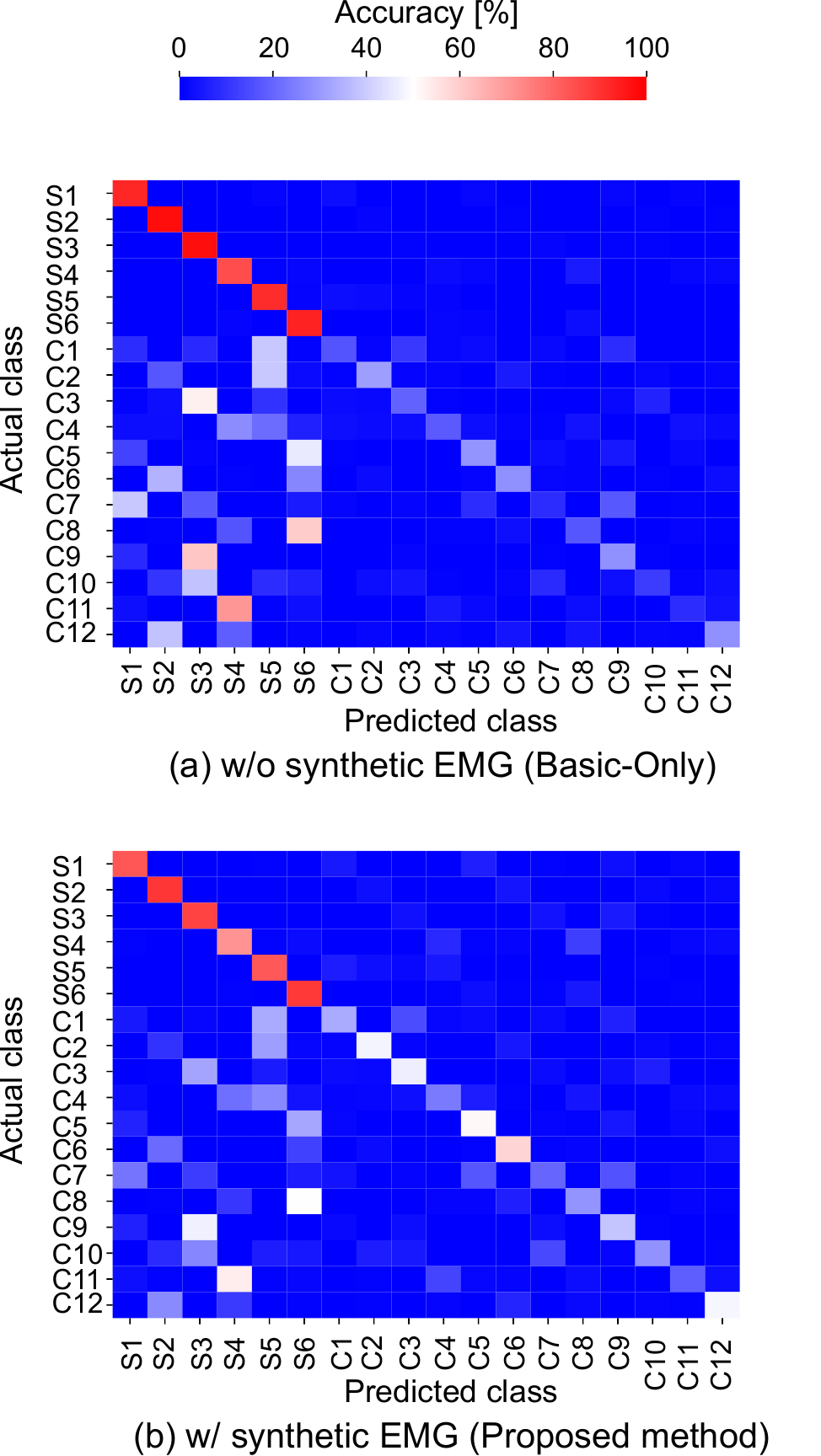}
	\caption{Confusion matrices comparing classification performance between (a) basic-only training and (b) proposed method with synthetic EMG data. Synthesis is performed at the input layer in the proposed method.}
	\label{confusion matrix}
\end{figure}

Fig.~\ref{confusion matrix} presents the averaged confusion matrices for the basic-only and proposed methods.
The colormap indicates higher accuracy with redder shades and lower accuracy with bluer shades.
The matrices demonstrate that while the basic-only method achieves high accuracy for basic motions, it fails to accurately classify most combined motions.
In contrast, the proposed method improves the accuracy for combined motions while largely maintaining the accuracy for basic motions.


\section{Discussion}

In this study, we aimed to achieve multi-DoF motion classification without utilizing actual combined motion data during training.
The generation of synthetic patterns at the input layer achieved higher classification accuracy compared to generation at hidden layers (Fig.~\ref{line plot}).
This finding suggests that linear synthesis produced better classification results than nonlinear synthesis for this particular task.
As shown in Fig.~\ref{pca}(a), synthetic patterns generated at the input layer exhibited features closely aligned with actual patterns.
In contrast, patterns generated at hidden layers deviated substantially from actual patterns (Fig.~\ref{pca}(b)).
These results indicate that as the synthetic data generation layer approached the final layer, the ability to represent intermediate values between basic motions diminished, resulting in decreased classification accuracy.

Based on these findings, we implemented synthetic data generation at the input layer in our proposed method.
The conventional method (basic-only), which focused solely on basic motion classification, demonstrated substantially lower accuracy for combined motions (Fig.~\ref{box plot}(b)).
This result suggests that models trained exclusively on basic motions could neither effectively capture the characteristics of combined motions nor distinguish between basic and combined motions.
In contrast, the proposed method improved classification accuracy for combined motions.
As shown in Fig.~\ref{radarchart}, the synthetic EMG patterns generated for combined motions exhibited similar characteristics to those of actual EMG patterns, enabling the model to effectively learn features associated with combined motions.

For basic motion classification, the basic-only method demonstrated superior performance compared to other methods (Fig.~\ref{box plot}(a)).
This higher accuracy can be attributed to its specialized training on basic motion data.
Notably, the proposed method achieved higher accuracy for basic motions than even the ideal fully-supervised method.
These results suggest that incorporating synthetic combined motion data into the training process enables robust classification of unseen combined motions while maintaining high accuracy for basic motions.

However, when considering the classification of all motions, a performance gap of approximately $30$\% persisted between the ideal fully-supervised method and the proposed method (Fig.~\ref{box plot}(c)).
As shown in Fig.~\ref{confusion matrix}, while the proposed method improved classification accuracy for motions that were identifiable using the conventional method, it still struggled with motions that were difficult to classify conventionally.
This limitation suggests that the synthetic EMG patterns did not fully replicate the characteristics of actual combined motions.
Indeed, as shown in Figs.~\ref{radarchart}(b), (d), and (f), the synthetic patterns in these specific cases exhibited clear deviations from actual patterns.
Although the overall pattern morphology was preserved, specific electrode values in the synthetic patterns showed disproportionate scaling compared to those in the actual pattern.
This discrepancy likely arose from complex pattern changes in actual combined motions, influenced by phenomena such as co-contraction~\cite{zajac2002biomechanics, zajac2003biomechanics, ramirez2023detection}.
The linear generation approach employed in our method could not fully capture these intricate features.

Furthermore, Nowak~\textit{et al.} demonstrated that optimal combination ratios for data generation may vary across different motions~\cite{nowak2016let}.
In the proposed method, we uniformly applied values sampled from a symmetric distribution with a mean of $0.5$ across all motions, potentially overlooking motion-specific asymmetries. 
This suggests that adapting the generation method to account for motion-specific characteristics could further improve classification accuracy.

\section{Conclusion}
This paper proposed a method for classifying unseen combined motions through synthetic EMG signal generation.
The proposed method generated synthetic EMG patterns for combined motion classes using measured patterns from basic motion classes.
By incorporating both actual and synthetic patterns into the training process, our method enables the classification of diverse motions with limited training data.

Experimental evaluation with eight healthy participants demonstrated that the proposed method improved classification accuracy for unseen combined motions in upper limb motion classification tasks.
Notably, synthesis at the input layer achieved higher accuracy compared to that at hidden layers.
However, the method showed limitations in synthetic data generation accuracy.
Specifically, the linear synthesis using convex combinations did not fully capture complex pattern changes, such as muscle co-contractions.

Future work will focus on improving synthetic data accuracy by mapping input data to a non-linear feature space, where basic motion patterns serve as bases, enabling more sophisticated data generation within this transformed space.
Additionally, we plan to develop methods for learning data generation patterns from a small number of participants and transferring these learned patterns to other subjects.

\addtolength{\textheight}{-12cm}


\bibliographystyle{IEEEtran}
\bibliography{EMBC2025}

\end{document}